# Equilibrium conformation of polymer chains with noncircular cross section

Shumin Zhao, Shengli Zhang, Zhenwei Yao, and Lei Zhang

*Department of Applied Physics, Xi'an Jiaotong University, Xi'an 710049, China*

In this paper we derive the general equilibrium equations of a polymer chain with a noncircular cross section by the variation of the free energy functional. From the equilibrium equation of the elastic ribbon we derive analytically the equilibrium conformations both of the helical ribbons and the twisted ribbons. We find that the pitch angle of the helical ribbons depends on the ratio of the torsional rigidity to the bending one. For the twisted ribbons, the rotation rate depends on the spontaneous torsion, which is determined by the elastic properties of the polymers. Our results for helical and twisted ribbons strongly indicate that the formation of these structures is determined by their elastic properties.

The equilibrium conformation of biopolymers and filaments is an important issue in molecular and cellular biology, and polymer physics [1]. For example, earlier theoretical models used elastic theory to describe the conformation of vital biomolecules such as proteins and DNA [2–6]. In general, such models of polymer chains involve the use of elastic coefficients, which describe the elastic properties of materials. However, up to now most models of polymer chains have only considered circular cross sections. The need to improve on these models by introducing the significant effects of geometry is obvious. For example, recently there have been numerous experimental observations of polymers with noncircular cross sections such as chemically defined lipid concentrate (CDLC) [7–10], gemini surfactants [11], synthetic nonracemic chiral polymers [12,13], and carbon nanotubes [14]. During the past decade there have been some theoretical attempts made to account for the effects of geometry in order to improve models. The helical structures of tilted chiral lipid bilayers were investigated by using a free energy model without considering the cross section [15]. The stretching instability of helical ribbons was discussed by using continuum phenomenological elastic models [16,17]. The transition between the helical and twisted ribbon structures of chiral materials has been studied both using continuum elastic theory and lattice Monte Carlo simulations [18,19]. For polymer chains with noncircular cross sections, the thermal fluctuations on the statistical properties of thin elastic filaments have also been investigated [20,21]. However, their equilibrium conformation equations have yet to be presented in the literature.

In this paper, we focus on polymers characterized by a nonzero thickness and a noncircular cross section. We present a general elastic model of polymer chains with noncircular cross sections and use it to discuss elastic ribbons. First we consider the free energy density as a general functional of the curvature, the torsion, and the twist angle of a polymer chain. Then by calculating the variation of the free energy functional, we obtain the equilibrium conformation equations of a polymer chain with a noncircular cross section. Finally we consider the case of an inextensible elastic ribbon and look at some special cases. This is used, in turn, to study the equilibrium conformations of the helical and twisted ribbons.

In general, a finite polymer chain with a noncircular cross section in three-dimensional (3D) space is modeled as an inextensible but deformable space curve parametrized by a contour length $s$ ($0 \leq s \leq L$, where $L$ is the total length of the polymer chain). The rotation of the cross section about the centerline is denoted by twist angle $\alpha(s)$. The free energy $\mathcal{F}$ of the polymer chain can be expressed as,

$$\mathcal{F} = \int F[\mathbf{x}(s), \alpha(s)] ds, \quad (1)$$

where $F$ is a scalar free energy density functional and depends on both the position vector $\mathbf{x}(s)$ and the twist angle $\alpha(s)$.

It is well known that the general configuration of a three-dimensional space curve $\mathbf{x}(s)$ can be described by an orthonormal triad of unit vectors $\{\mathbf{t}_i(s)\}$, where $\mathbf{t}_1$ and $\mathbf{t}_2$ are oriented along the principal axes of the cross section and $\mathbf{t}_3 = d\mathbf{x}/ds$ is tangent to the curve at point $s$. The vectors $\mathbf{t}_i$ satisfy the generalized Frenet equations, $d\mathbf{t}_i(s)/ds = -\sum_{j,k} \epsilon_{ijk} \omega_j(s) \mathbf{t}_k(s)$, where $\epsilon_{ijk}$ is the antisymmetric tensor and $\{\omega_i(s)\}$ are the generalized torsion parameters [21]. In general, the generalized torsions $\{\omega_i(s)\}$ are determined by the curvature $k$, torsion $\tau$, and the angle $\alpha$, as $\omega_1 = k \cos \alpha$, $\omega_2 = k \sin \alpha$, and $\omega_3 = \tau + d\alpha/ds$ [21]. The general free energy $\mathcal{F}$ has the form

$$\mathcal{F} = \int F[\kappa(s), \tau(s), \alpha(s), \alpha_s(s)] ds, \quad (2)$$

where $\alpha_s(s) = d\alpha(s)/ds$ is the rate of rotation of the cross section along the centerline of the polymer chain. In other words, the two neighboring cross sections at a distance $ds$ rotate by a relative angle $d\alpha(s) = \alpha_s(s) ds$. The functional $F$ will in general involve elastic constants, depending both on the elasticity of materials and the geometrical shape of the cross section.

Equation (2) is a general expression of the elastic energy of polymer chains with noncircular cross sections, which is also valid for elastic filaments [22]. Taking the variation $\delta \mathcal{F} = 0$, one can derive the equilibrium conformation equations of the polymer chains. The variation of the space form of the polymer chains can be written as follows:

$$\delta \mathcal{F} = \delta F_1 + \delta F_2 + \delta F_3 + \delta F_4 + \delta F_5, \quad (3)$$

in which $\delta F_1 = \int F_1 \delta \kappa(s) ds$, $\delta F_2 = \int F_2 \delta \tau(s) ds$, $\delta F_3 = \int F_3 \delta \alpha(s) ds$, $\delta F_4 = \int F_4 \delta \alpha_s(s) ds$, $\delta F_5 = \int F \delta ds$, and $F_1$

$= \partial F/\partial\kappa$, $F_2 = \partial F/\partial\tau$, $F_3 = \partial F/\partial\alpha$, $F_4 = \partial F/\partial\alpha_s$.

In order to give a more detailed expression of $\delta\mathcal{F}$, we must determine the variations $\delta\kappa(s)$, $\delta\tau(s)$, $\delta\alpha(s)$, and $\delta\alpha_s(s)$. The variations $\delta\kappa(s)$ and $\delta\tau(s)$ were obtained previously by using differential geometry methods [22,23]. Similarly, the variation of the twist angle $\delta\alpha(s)$ is obtained as follows:

$$\delta\alpha_s(s) = \frac{d}{ds}\delta\alpha(s) - d\alpha(s)\frac{\delta ds}{ds^2} = \varepsilon_4' - \alpha_s(\varepsilon_1 - \varepsilon_2). \quad (4)$$

where $\varepsilon_i(s)$ are the variations of the space form of the polymer chain $[\delta\mathbf{x}(s) = \varepsilon_i(s)\mathbf{e}_i(s), i=1,2,3$, with the orthonormal Frenet basis $\{\mathbf{e}_i\}]$ and the variation of the twist angle $\delta\alpha(s)$, $\varepsilon_4$, is assumed to be a small quantity.

For the simplest case of a polymer chain which is inextensible, the terms in $\delta\mathcal{F}$ do not depend on the variation $\mathcal{E}_1(s)$ along the tangential direction of the curve $\mathbf{x}(s)$. Using Eq. (4) and the Eqs. (2.20) and (2.27) in Ref. [5], we can derive the variation of the free energy from Eq. (3),

$$\delta F = \int ds \left( \left\{ \frac{d^2}{ds^2}\left(\frac{2F_2\tau}{\kappa} + F_1\right) + \frac{d}{ds}\left(\frac{2F_2\kappa'\tau}{\kappa^2} - \frac{3F_2\tau'}{\kappa}\right) \right. \right.$$
$$+ \left[ F_1(\kappa^2 - \tau^2) + F_2\left(2\kappa\tau - \frac{\kappa'\tau'}{\kappa^2} + \frac{\tau''}{\kappa}\right) \right] - F\kappa$$
$$+ \alpha_s\kappa F_4 \right\}\varepsilon_2 + \left\{ \frac{d^3}{ds^3}\left(\frac{F_2}{\kappa}\right) - \frac{d^2}{ds^2}\left(\frac{F_2\kappa'}{\kappa^2}\right) - \frac{d}{ds}\left[\frac{F_2}{\kappa}(\kappa^2 - \tau^2) - 2\tau F_1\right] - F_1\tau' + F_2\kappa' - \frac{2F\tau\tau'}{\kappa} \right\}\varepsilon_3$$
$$+ \left\{ F_3 - \frac{d}{ds}F_4 \right\}\varepsilon_4 \right) = 0. \quad (5)$$

The equilibrium conformations of the polymer chains are determined by Eq. (5), which leads to the following three conditions:

$$\frac{d^2}{ds^2}\left(\frac{2F_2\tau}{\kappa} + F_1\right) + \frac{d}{ds}\left(\frac{2F_2\kappa'\tau}{\kappa^2} - \frac{3F_2\tau'}{\kappa}\right) + \left[ F_1(\kappa^2 - \tau^2) \right.$$
$$+ F_2\left(2\kappa\tau - \frac{\kappa'\tau'}{\kappa^2} + \frac{\tau''}{\kappa}\right) \right] - F\kappa + \alpha_s\kappa F_4 = 0, \quad (6)$$

$$\frac{d^3}{ds^3}\left(\frac{F_2}{\kappa}\right) - \frac{d^2}{ds^2}\left(\frac{F_2\kappa'}{\kappa^2}\right) - \frac{d}{ds}\left[\frac{F_2}{\kappa}(\kappa^2 - \tau^2) - 2\tau F_1\right] - F_1\tau'$$
$$+ F_2\kappa' - \frac{2F\tau\tau'}{\kappa} = 0, \quad (7)$$

$$F_3 - \frac{d}{ds}F_4 = 0. \quad (8)$$

These conformation equations provide a general description for the equilibrium conformations of the polymer chains with noncircular cross sections.

Next we apply these results to a particular case—the elastic ribbon. The elastic ribbon is modeled as an inextensible ribbon whose elastic properties are characterized by the elastic coefficients, including the bending rigidities $A$ and $B$, and the torsional rigidity $C$. Following J. F. Marko [24] and Y. Rabin et al. [21], the corresponding free energy density of this elastic ribbon is,

$$F = F(\kappa, \tau, \alpha, \alpha_s) = \frac{A}{2}\kappa^2 \cos\alpha + \frac{B}{2}\kappa^2 \sin\alpha + C(\tau + \alpha_s - \tau_0)^2, \quad (9)$$

where $\tau_0$ is the spontaneous torsion. We work in units with the temperature $k_BT=1$, so that $A$, $B$, and $C$ have dimensions of length. In general, the coefficients $A$, $B$, and $C$ depend on the elastic properties of materials and the geometry of the cross section. For isotropic materials, the rigidities can be expressed in terms of the shear modulus $\mu$, Young's modulus $E$, and the principal moments of inertia, which depend on the shape of the cross section [25].

Using Eq. (9) we can investigate the equilibrium conformation of the elastic ribbon. As a simple but illustrative example, we consider the case in which the curvature and torsion of the curve $\mathbf{x}(s)$ does not depend on $s$. In order to find the equilibrium conformation of this ribbon, one first derives the corresponding equilibrium equations as

$$\frac{\kappa}{2}\{(A+B)\kappa^2 - 2(A+B-3C)\tau^2 + (A-B)(\kappa^2 - 2\tau^2)\cos 2\alpha$$
$$- 4C\tau\tau_0 - 2C\tau_0^2 + 8C\tau\alpha' + 2[C + 4(B-A)\cos 2\alpha]\alpha'^2$$
$$+ 4(B-A)\alpha'' \sin 2\alpha\} + \frac{4}{\kappa}C\tau\alpha^{(3)} = 0, \quad (10)$$

$$4(A-B)\kappa\tau\alpha' \sin 2\alpha + \frac{2C}{\kappa}[(\kappa^2 - \tau^2)\alpha'' - \alpha^{(4)}] = 0, \quad (11)$$

$$(B-A)\kappa^2 \sin 2\alpha - 2C\alpha'' = 0. \quad (12)$$

The equilibrium conformation of the typical polymer chains with noncircular cross sections is then determined by Eqs. (10)–(12). It is found that two types of ribbons satisfy Eqs. (10)–(12): helical and twisted, each of which will be discussed in detail below.

*Helical ribbons*. In order to reduce the number of variables, we consider a ribbon with a highly asymmetric cross section, $B/A \rightarrow \infty$, in which case the ribbon becomes lamelliform and the twist angle of the cross section $\alpha(s) = 0$ [17]. Thus the equilibrium conformation of the elastic ribbon is completely determined by an $s$-independent curvature $\kappa$ and torsion $\tau$. Since a helix is characterized by its radius $r$ and pitch $2\pi h$, substituting $\kappa = r/(r^2 + h^2)$ and $\tau = h/(r^2 + h^2)$ into Eq. (12), we obtain

$$3Ch^2 + A(r^2 - 2h^2) - C\tau_0(r^2 + h^2)[2h + (r^2 + h^2)\tau_0] = 0, \quad (13)$$

which is the optimal conformation equation for helical ribbons.

It is seen from Eq. (13) that the conformation of these helical ribbons depends on the elastic properties of the helical ribbons. For a given helical ribbon, the elastic parameters $A$ and $C$ are fixed. When the spontaneous torsion $\tau_0 = 0$, we

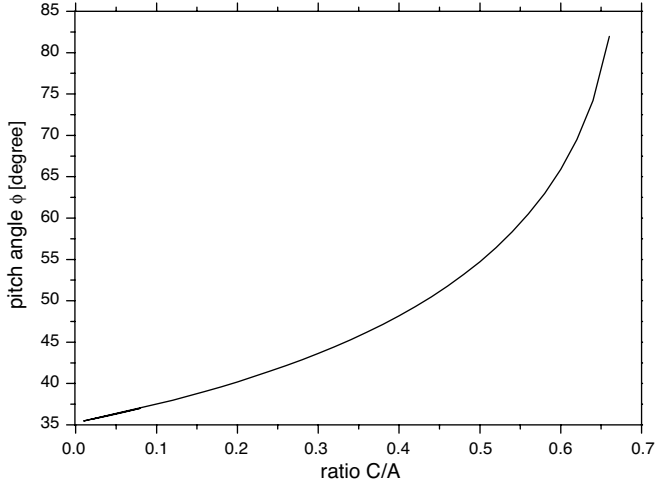

FIG. 1. The pitch angle of helices as a function of the ratio $C/A$.

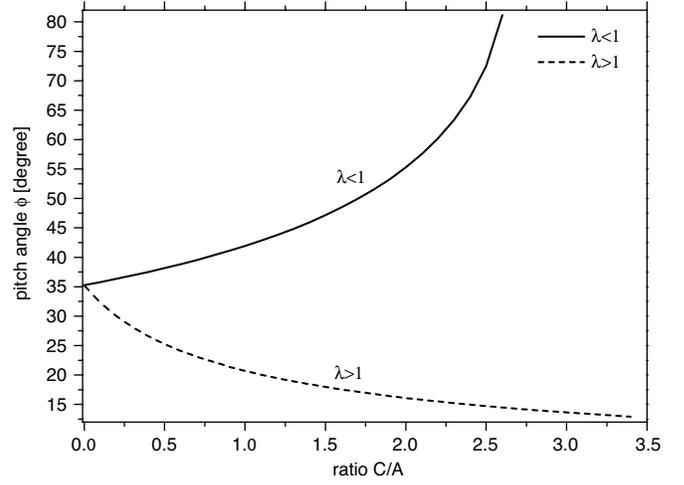

FIG. 2. The pitch angle $\phi$ versus the ratio $C/A$ for $\tau_0/\tau=0.8$ (solid curve) and $\tau_0/\tau=2$ (dashed curve).

can obtain $r/h=\kappa/\tau=[(2A-3C)/A]^{1/2}$ and the pitch angle as

$$\phi = \arctan[1/(2-3C/A)^{1/2}]. \quad (14)$$

This expression shows that the value of the pitch angle of the helices depends on the ratio of the torsional rigidity $C$ to bending rigidity $A$. From Eq. (14) we can get the critical value of the ratio $C/A=2/3$. When $C/A>2/3$ there should be no helical ribbons without spontaneous torsion. Figure 1 shows the results when the ratio $C/A$ is in the range 0 to 2/3. We find that the value of the pitch angle increases from 35° to 70° as $C/A$ increases from 0 to 2/3.

Now we consider the helical ribbons with spontaneous torsion $\tau_0 \neq 0$. This time the pitch angle is

$$\phi = \arctan\{1/[2+[(1+\lambda)^2-4]C/A]^{1/2}\}, \quad (15)$$

where the parameter $\lambda = \tau_0/\tau$. In Fig. 2 we present plots of the pitch angle $\phi$ as a function of the ratio $C/A$ for fixed values of $\lambda$. We see that the pitch angle decreases as the ratio $C/A$ increases when $\lambda$ is greater than 1. The actual torsion is larger than the spontaneous torsion in this case. On the contrary, the pitch angle increases as the ratio $C/A$ increases for $\lambda$ less than 1. In that case, the actual torsion is smaller than the spontaneous torsion.

Figure 3 gives the change of helical ribbons from low pitch to high pitch. Each structure is characterized by the pitch angle $\phi$. For three values of the pitch angle $\phi=13°$, 40°, and 50°, we find, respectively, $C/A=1.85$, 0.23, and 0.5 (in this analysis we have: $\lambda=3$ for $\phi=13$, and $\lambda=0.3$ for $\phi=40°$ and 50°). These results show that the conformation of the helical ribbons varies with the elastic properties of the ribbons.

In particular, in elastic models of DNA, the torsional rigidity $C$ is commonly assumed to exceed the bending rigidity $A$ [26]. In the case of large torsional rigidity ($C/A \to \infty$) the torsion $\tau = \tau_0$, and the curvature obeys the relation $\kappa^2 = 2\tau^2$, from our calculations it is easy to compute $r/h=\sqrt{2}\simeq 1.41$, $\phi \simeq 35°$, which agree well with Z-DNA [27] (with $p=4.46$ nm and $r=0.90$ nm, thus $r/h=2\pi r/p \simeq 1.27$).

For isotropic ribbons we have the ratio $C/A \simeq 4\mu/E = 2/(1+\sigma)$. Then Eq. (15) becomes

$$\phi = \arctan\{1/[2+2[(1+\lambda)^2-4]/(1+\sigma)]^{1/2}\}, \quad (16)$$

where $\sigma$ is Poisson's ratio in the range $-1 \leq \sigma \leq 1/2$ [25]. Solutions to Eq. (16) are real only when $\lambda > \sqrt{5/2}-1$, which implies that the isotropic helical ribbons with $\lambda < \sqrt{5/2}-1$ are unstable. Similarly, $\phi$ decreases with increasing $\sigma$ when $\sqrt{5/2}-1 < \lambda < 1$, while it increases with increasing $\sigma$ when $\lambda > 1$. Furthermore, results for incompressible materials coincide with the critical value $C/A=4/3$ [17], which means the ratio $C/A$ can be increased beyond this critical value by reducing $\sigma$ or by using anisotropic materials with high resistance to twist.

*Twisted ribbons*. For vanishing curvature and torsion ($\kappa=0$, $\tau=0$), Eqs. (10) and (11) obviously become identical. From Eq. (12) we have $C\alpha''=0$, i.e., $\alpha'(s)=$const. Thus the rotation rate is constant along the ribbon. The total angle of rotation of the upper end of the ribbon relative to the lower end is $\alpha'L$.

In Eq. (9), by minimizing the free energy of the ribbon

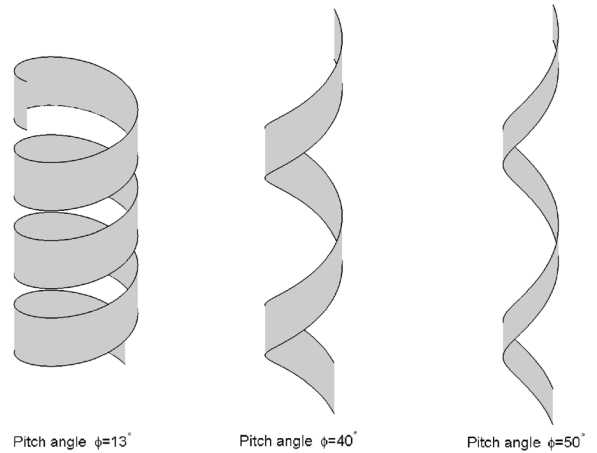

FIG. 3. Series of helical ribbons with the parameter $\phi=13°$, 40°, and 50°.

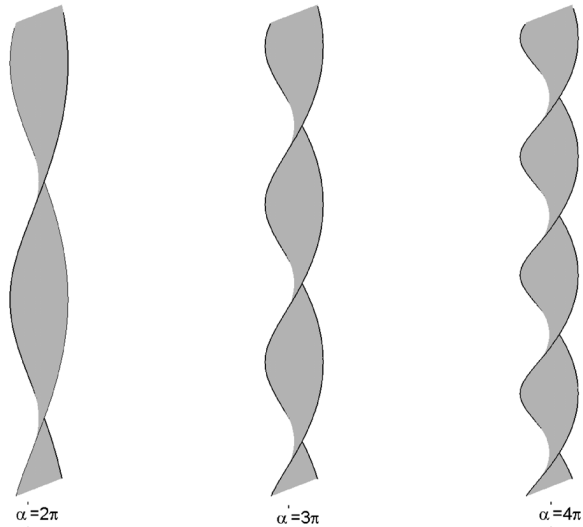

FIG. 4. Series of twisted ribbons with the parameter $\tau_0 = 2\pi$, $3\pi$, and $4\pi$.

with vanishing curvature and torsion we obtain the following expression

$$\alpha'(s) = \tau_0. \quad (17)$$

It corresponds to twisted ribbons with rotation rate $\tau_0$. We see that the spontaneous torsion is just the angle of rotation per unit length of the ribbons. In general the spontaneous torsion $\tau_0$ depends on the elastic properties of twisted ribbons. If $\tau_0$ is large enough, the free energy of twisted ribbons is lower than that of straight ones.

We believe that twist occurs mainly because the spontaneous torsion $\tau_0$ depends on the elastic properties. Figure 4 illustrates the change of twisted ribbons from low twist pitch to high twist pitch with $\tau_0 = 2\pi$, $3\pi$, and $4\pi$. These results show that the conformation of the twisted ribbons varies with the different elastic properties of the ribbons, which provides a possible explanation for the different structures of twisted ribbons.

In conclusion, we have derived the general equilibrium conformation equations of polymer chains with noncircular cross sections. For the elastic ribbon model, we obtain two solutions and show that the equilibrium conformation varies with the different elastic properties of the ribbons. It is found that the pitch angle depends on the ratio of the bending to the torsional rigidity of the helical ribbons. The twist pitch of the twisted ribbons strongly depends on the spontaneous torsion $\tau_0$, which also depends on the elastic properties of the ribbons. Thus, the known elastic properties of biopolymers such as DNA and proteins may be very helpful in understanding their formation mechanisms. Our theory can be verified by measuring the structures and corresponding elastic properties of biopolymers experimentally.

We thank Professor Y. Rabin for drawing our attention to the polymer chains with noncircular cross sections. This work was supported by NNSF of China under Grants No. 10374075 and No. 10547135.